\documentclass[twoside,fleqn]{ActaStyle}
\usepackage{times,cite}

\usepackage{lscape}             % if you need landscape enviroinment
\usepackage{graphicx,epsfig}    % if you need to include figures in PostScript
\usepackage[tbtags]{amsmath}    % if you need more math
\def\authorheading{H.~Machner}

\def\shorttitle{$\eta$ Meson Physics at GEM}

\begin{document}
%% ---------------------------------------------------

%%%  page range, first and last page
\pagerange{1}{10}

%%% paper title
\title{%
$\eta$ MESON PHYSICS AT GEM. }

%%% author(s) and address(es)
\author{%  author(s)
H.~Machner\email{h.machner@fz-juelich.de }\\representing the GEM collaboration\\
M.~Abdel-Bary$^{a}$, A.~Budzanowski$^{c}$, A.~Chatterjee$^{g}$,
J.~Ernst$^{f}$, P.~Hawranek$^{a,b}$, R.~Jahn$^{f}$, V.~Jha$^{g}$,
K.~Kilian$^{a}$, S.~Kliczewski$^{c}$, Da.~Kirillov$^{a}$,
Di.~Kirillov$^{k}$, D.~Kolev$^{e}$, M.~Kravcikova$^{j}$,
T.~Kutsarova$^{d}$, M.~Lesiak$^{a,b}$, J.~Lieb$^{h}$,
H.~Machner$^{a,n}$ A.~Magiera$^{b}$, R.~Maier$^{a}$,
G.~Martinska$^{i}$, S.~Nedev$^{l}$, N.~Piskunov$^{k}$,
D.~Prasuhn$^{a}$, D.~Proti\'c$^{a}$, P.~von Rossen$^{a}$,
B.~J.~Roy$^{g}$, I.~Sitnik$^{k}$, R.~Siudak$^{c,f}$,
R.~Tsenov$^{e}$, M.~Ulicny$^{i}$, J.~Urban$^{i}$,
G.~Vankova$^{a,e}$, C.~Wilkin$^{m}$ }
%%%%%%%%%%%%%%%%%%%%%%%%%%%%%%%%%%%%%%%%%%%%%%%%%%%%%%%%%%%%%%%%%%%%%%%%%%%%%%%%%%%%%%%%%%%
{%  address(es)
$^a${Institut f\"{u}r Kernphysik, Forschungszentrum J\"{u}lich, 52425
J\"{u}lich, Germany}\\
$^b${Institute of Physics, Jagellonian University, Krakow, Poland}\\
$^c${Institute of Nuclear Physics, Polish Academy of Sciences,
Krakow, Poland}\\
$^d${Institute of Nuclear Physics and Nuclear Energy, Sofia,
Bulgaria}\\
$^e${Physics Faculty, University of Sofia, Sofia, Bulgaria}\\
$^f${Helmholtz-Institut f\"{u}r Strahlen- und Kernphysik der Universit\"{a}t
Bonn, Bonn, Germany}\\
$^g${Nuclear Physics Division, BARC, Bombay, India}\\
$^h${Physics Department, George Mason University, Fairfax, Virginia,
USA}\\
$^i${P.~J.~Safarik University, Kosice, Slovakia}\\
$^j${Technical University, Kosice, Kosice, Slovakia}\\
$^k${Laboratory for High Energies, JINR Dubna, Russia}\\
$^l${University of Chemical Technology and Metallurgy, Sofia,
Bulgaria}\\
$^m${Department of Physics \& Astronomy, UCL, London, U.K.}\\
$^n${Fachbereich Physik, University Duisburg-Essen}}

%%%%%%%%%%%%%%%%%%%%%%%%%%%%%%%%%%%%%%%%%%%%%%%%%%%%%%%%%%%%%%%%%%%%%%%%%%%%%%%%%%%%%%%%%%%

%%% Date of submition
\day{20. Oct. 2005}

%%% abstract of the paper
\abstract{%
Some experimental studies of $\eta$ production and $\eta$
interactions performed or presently under way by the GEM
collaboration at COSY J\"{u}lich are reviewed.}

%%% PASC numbers of your article

% %%%%%%%%%%%%%%%%%%%%%%%%%%%%%%%%%%%%%%%%%%%%%%%%%%%%%%%%%%%%%%%%%
\section{Introduction}
\begin{figure}
\includegraphics[height=5.5 cm]{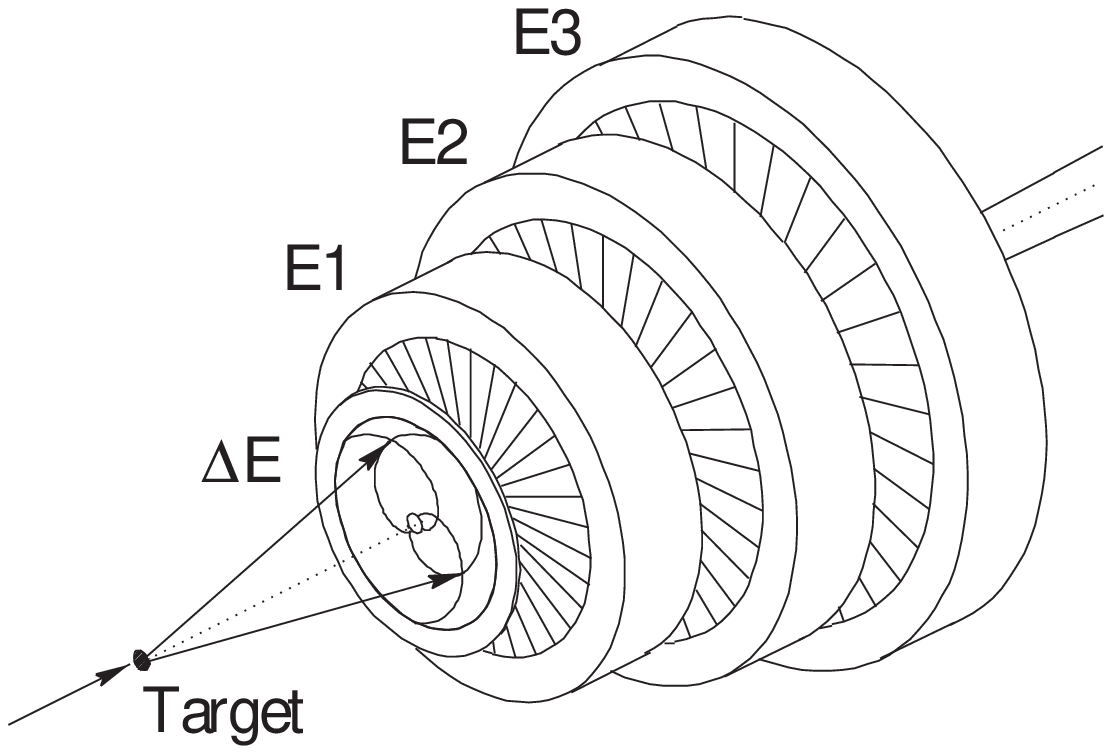}
\includegraphics[height=5.5 cm]{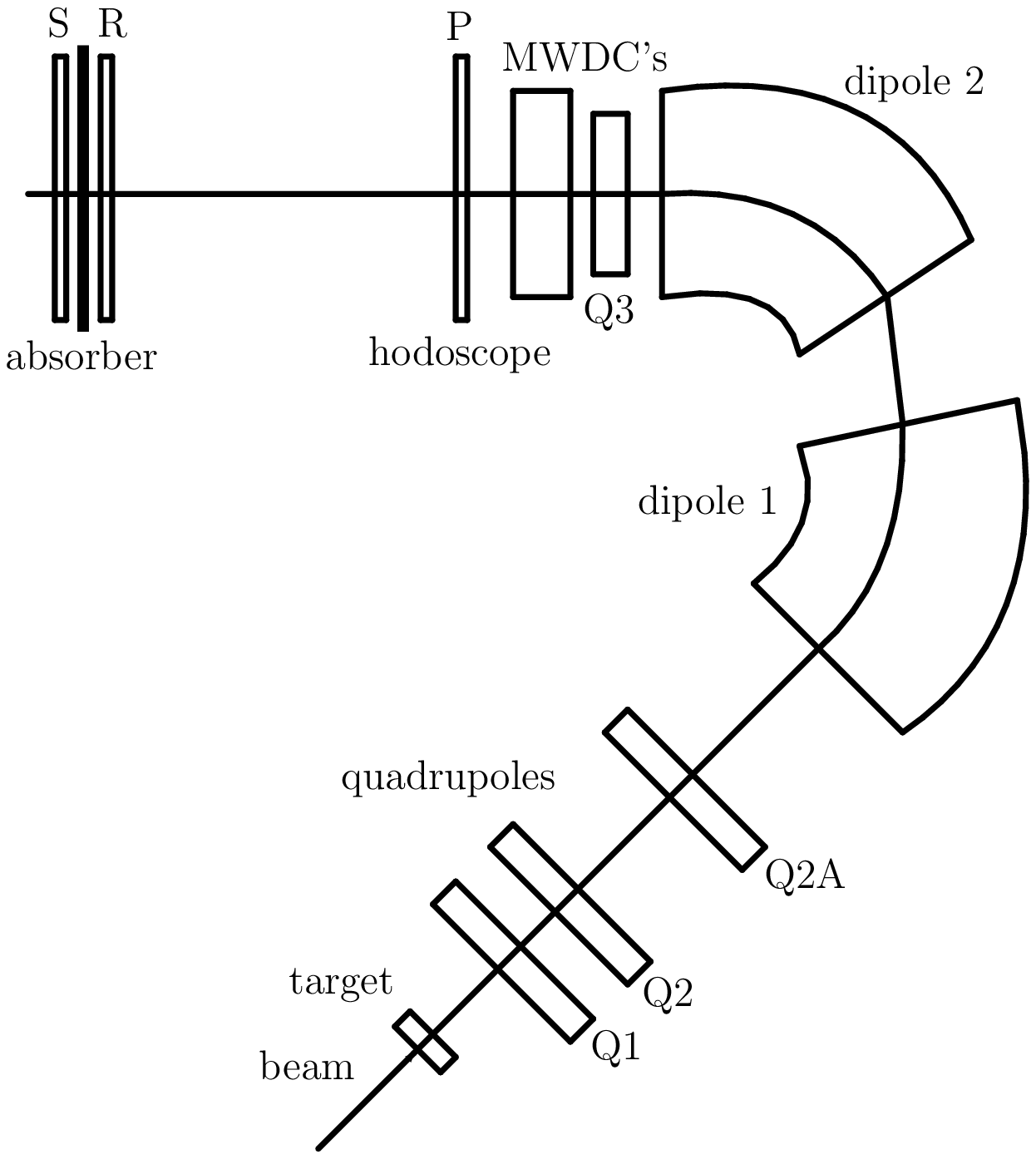}
\caption{Left: Perspective view of the germanium wall. An event with
two hits is indicated. The diameter of the active area on last diode
is 7.7 cm. Right: Cross section of the magnetic spectrograph Big
Karl. Note the differences in scale of the two drawings. The flight
path from the target to the focal plane is $\approx 15 $ m long.}
\label{fig:GEM}
\end{figure}

% %%%%%%%%%%%%%%%%%%%%%%%%%%%%%%%%%%%%%%%%%%%%%%%%%%%%%%%%%%%%%%%%%
\label{sec:intr} \setcounter{section}{1}\setcounter{equation}{0}
%%%%%%%%%%%%%%%%%%%%%%%%%%%%%%%%%%%%%%%%%%%%%%%%%%%%%%%%%%%%%%%%%%%
The collaboration operates the GEM detector, which is a combination
of a stack of \emph{GE}rmanium diodes and a \emph{M}agnetic
spectrograph. The germanium wall \cite{Betigeri99} consists of four
annular detectors. The first one is position sensitive with 200
Archimedes spirals on the front and also on the rear side but with
opposite orientation. In this way 40000 pixels are defined. It
provides position (see Fig. \ref{fig:GEM}) and $\Delta E$
information. The thick detectors that follow are segmented into 32
wedges. The total thickness of the germanium wall is $\approx 51$
mm. Due to the different topologies one can identify multiple hits.
The magnetic spectrograph is schematically also shown in Fig.
\ref{fig:GEM}. It is a high resolution device in which reaction
products pass through three quadrupole magnets and two dipole
magnets.  It has point to parallel imaging in the vertical and point
to point imaging in the horizontal direction. The last quadrupole
magnet $Q3$ is not in use in this operation mode. The direction of
the reaction products is measured with MWDC's, twelve layers in two
packs. They are followed by scintillator hodoscopes $P$, $R$ and $S$
which give $\Delta E$ information and allow for a time of flight
(TOF) measurement. For further particle identification absorber
material can be placed between the last two layers. Additional
details of the detector  are given in \cite{Bojowald02}.
For a search for the existence of bound $\eta$-nuclear states an
additional detector ENSTAR was added surrounding the target. It was
recently used in a search employing the reaction $p+{^{27}}Al\to
{^3He}+(_\eta{^{25}Mg})$ at recoil-free conditions followed by a
second step $\eta + n\to N^{0*}\to \pi^-+p$. The ${^3He}$ was
detected in the magnetic spectrograph while the second step was
identified in ENSTAR. Details are discussed in the contribution by
Gillitzer \cite{Gillitzer05}.

%%%%%%%%%%%%%%%%%%%%%%%%%%%%%%%%%%%%%%%%%%%%%%%%%%%%%%%%%%%%%%%%%%%
\section{The reaction $p+d\to {^3He}+\eta$}\label{sec:reaction_pd}
\begin{figure}
\begin{center}
\includegraphics[width=8 cm]{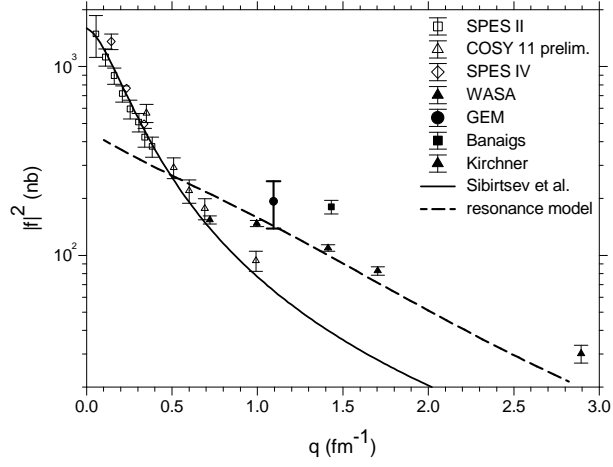}
\caption{The spin averaged matrix element for the reaction
$pd\to{^3He}\eta$ as function of the transferred momentum. The data
are from \cite{Mayer96} (open squares), \cite{Khoukaz04a} (open
triangles), \cite{Berger88} (rhombs), \cite{Bilger02} (full
triangle), \cite{Betigeri00} (full dot), \cite{Banaigs73} (full
squares) and \cite{Kirchner93} (full triangle). The solid curve is a
fit \cite{Sibirtsev05} and the dashed curve the resononce model
calculation \cite{Betigeri00}.} \label{fig:f_ps2eta}
\end{center}
\end{figure}

%%%%%%%%%%%%%%%%%%%%%%%%%%%%%%%%%%%%%%%%%%%%%%%%%%%%%%%%%%%%%%%%%%%
The reaction $p+d\to {^3He}+\eta$ is of interest to study the
$\eta$-nucleus scattering length.  In Fig. \ref{fig:f_ps2eta} we
compare the measurements of different groups on the value of the
spin-averaged matrix element
\begin{equation}
|f|^2=\frac{\sigma _{tot}}{4\pi} \frac{p_p}{p_\eta}.
\end{equation}
as function of the transferred momentum $q=p_p-p_\eta$.
The close to threshold data are from various groups at SATURNE
\cite{Banaigs73, Berger88, Mayer96, Kirchner93} as well as more
recent data from GEM \cite{Betigeri00}, WASA \cite{Bilger02} and
COSY11 \cite{Khoukaz04b}. These latter data are not in agreement
with each other. This is especially true if in addition angular
distributions are compared. Obviously more insight is necessary to
clarify the situation. This may come from the newer measurement from
COSY11 with inverse kinematics where the detector has a larger
acceptance \cite{Smyrski05}. The solid curve is a fit to the the
data assuming $s$-wave production and final state interactions
 \cite{Sibirtsev05}. The dashed curve assumes
the reaction to proceed via a resonance. The matrix element is a
Breit-Wigner form
% MathType!MTEF!2!1!+-
% feaaeaart1ev0aaatCvAUfeBSjuyZL2yd9gzLbvyNv2CaerbuLwBLn
% hiov2DGi1BTfMBaeXatLxBI9gBaerbd9wDYLwzYbItLDharqqtubsr
% 4rNCHbGeaGqiVCI8FfYJH8sipiYdHaVhbbf9v8qqaqFr0xc9pk0xbb
% a9q8WqFfeaY-biLkVcLq-JHqpepeea0-as0Fb9pgeaYRXxe9vr0-vr
% 0-vqpWqaaeaabaGaaiaacaqaaeaadaqaaqaaaOqaaiaacYhacaWGMb
% GaaiiFamaaCaaaleqabaGaaGOmaaaakiabg2da9maalaaabaGaamyq
% aiabfo5ahnaaDaaaleaacaWGYbaabaGaaGOmaaaaaOqaaiaacIcada
% GcaaqaaiaadohaaSqabaGccqGHsisldaGcaaqaaiaadohadaWgaaWc
% baGaamOCaaqabaaabeaakiaacMcadaahaaWcbeqaaiaaikdaaaGccq
% GHRaWkcqqHtoWrcaGGOaWaaOaaaeaacaWGZbaaleqaaOGaaiykamaa
% CaaaleqabaGaaGOmaaaaaaaaaa!4A43!
\begin{equation}\label{equ:Breit-Wigner}
|f|^2  = \frac{{A\Gamma _r^2 }} {{(\sqrt s  - \sqrt {s_r } )^2  +
\Gamma (\sqrt s )^2 }}
\end{equation}
with a momentum dependent width
% MathType!MTEF!2!1!+-
% feaaeaart1ev0aaatCvAUfeBSjuyZL2yd9gzLbvyNv2CaerbuLwBLn
% hiov2DGi1BTfMBaeXatLxBI9gBaerbd9wDYLwzYbItLDharqqtubsr
% 4rNCHbGeaGqiVu0Je9sqqrpepC0xbbL8F4rqqrFfpeea0xe9Lq-Jc9
% vqaqpepm0xbba9pwe9Q8fs0-yqaqpepae9pg0FirpepeKkFr0xfr-x
% fr-xb9adbaqaaeGaciGaaiaabeqaamaabaabaaGcbaGaeu4KdCKaai
% ikamaakaaabaGaam4CaaWcbeaakiaacMcacqGH9aqpcqqHtoWrdaWg
% aaWcbaGaamOCaaqabaGccaGGOaGaamOyamaaBaaaleaacqaH3oaAae
% qaaOWaaSaaaeaacaWGWbWaa0baaSqaaiabeE7aObqaaiaacQcaaaaa
% keaacaWGWbWaa0baaSqaaiabeE7aOjaacYcacaWGYbaabaGaaiOkaa
% aaaaGccqGHRaWkcaWGIbWaaSbaaSqaaiabec8aWbqabaGcdaWcaaqa
% aiaadchadaqhaaWcbaGaeqiWdahabaGaaiOkaaaaaOqaaiaadchada
% qhaaWcbaGaeqiWdaNaaiilaiaadkhaaeaacaGGQaaaaaaakiabgUca
% RiaadkgadaWgaaWcbaGaeqiWdaNaeqiWdahabeaakiaacMcaaaa!5C61!
\begin{equation}\label{equ:width}
\Gamma (\sqrt s ) = \Gamma _r (b_\eta  \frac{{p_\eta ^* }} {{p_{\eta
,r}^* }} + b_\pi  \frac{{p_\pi ^* }} {{p_{\pi ,r}^* }} + b_{\pi \pi
} ).
\end{equation}
$\Gamma_r$ is the width at the resonance $\sqrt{s_r}$. As in
photoproduction \cite{Krusche96} we assumed $\sqrt{s_r}=1540$ MeV
and $\Gamma _r=200$ MeV. The branching ratios $b_i$ were taken from
the particle data group (PDG) \cite{PDG}. The absolute normalization
is arbitrary.

\section{The reaction $\vec d+d\to \eta
+\alpha$}\label{sec:reaction_dd}

\begin{figure}[h]\centering
\includegraphics[width=7 cm]{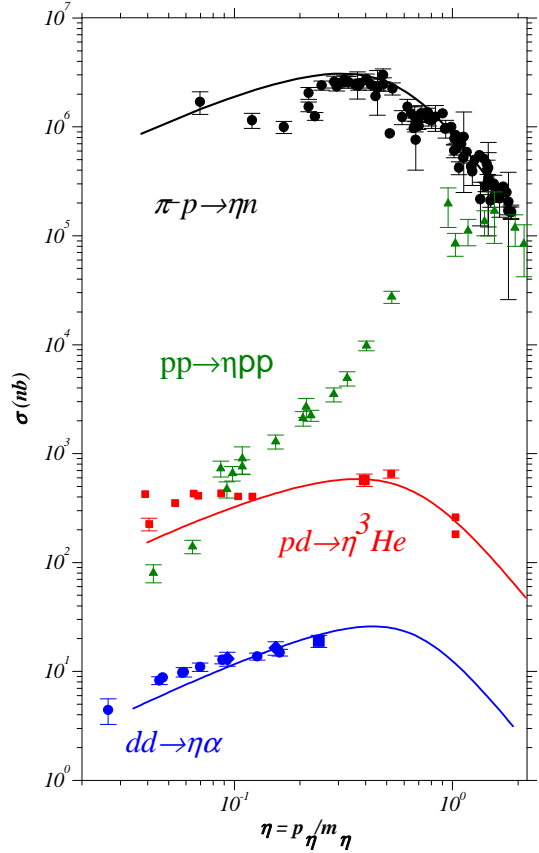}
\caption{Excitation functions for the indicated reactions. Points
from GEM are always indicated by squares, the new data from ANKE
\cite{Wronska05} as rhombi. Data for the reaction $pd\to {^3He}\eta$
from Refs. \cite{Bilger02} and \cite{Khoukaz04a} have been omitted.
The solid curves are the predictions of the resonance model
discussed in the text.} \label{Fig:exfu_all}
\end{figure}

Similar to the case of the previous reaction, the study of this
reaction is driven by the question whether a strong bound
$\eta$-nucleus system exists. Theory predicts that a heavier system
should result in stronger binding. Close to threshold only total
cross sections exist so far \cite{Frascaria94, Willis97} and only
recently the first angular distributions become available
\cite{Wronska05}. Since they will be discussed within this meeting
\cite{Wronska05a} we will concentrate on the energy dependence of
the total cross section. Also GEM has a preliminary value for the
total cross section measured at a beam momentum of 2.39 GeV/c. All
the known points are included in Fig. \ref{Fig:exfu_all}.
It is interesting to note that the new data follow the resonance
model prediction (Eq. (\ref{equ:Breit-Wigner}) together with Eq.
(\ref{equ:width})) which was previously adjusted to the data from
Willis et al. \cite{Willis97}. It also accounts for the $\pi^-p\to
\eta n$ cross sections, however, the excitation curve for $pp\to
\eta pp$ shows a completely different behavior. We can summarize the
present observations that reactions with two particles in the final
state seem to have a quite similar behavior that is different from
reactions with three particles in the final state.  The $dd\to
\eta\alpha$ reaction allows the extraction of the real and imaginary
parts of the partial wave amplitudes if one measures the vector and
tensor analyzing powers in addition to the differential cross
section. If we assume that there is only s- and p-wave in the
initial state, the polarized differential cross section for
transversely  polarized deuterons is given by
% MathType!MTEF!2!1!+-
% feaaeaart1ev0aaatCvAUfeBSjuyZL2yd9gzLbvyNv2CaerbuLwBLn
% hiov2DGi1BTfMBaeXatLxBI9gBaerbd9wDYLwzYbItLDharqqtubsr
% 4rNCHbGeaGqiVu0Je9sqqrpepC0xbbL8F4rqqrFfpeea0xe9Lq-Jc9
% vqaqpepm0xbba9pwe9Q8fs0-yqaqpepae9pg0FirpepeKkFr0xfr-x
% fr-xb9adbaqaaeGaciGaaiaabeqaamaabaabaaGcbaWaaeWaaeaada
% WcaaqaaiaadsgacqGHdpWCaeaacaWGKbGaeyyQdCfaamaabmaabaGa
% eqiUdeNaaiilaiabgw9aQbGaayjkaiaawMcaaaGaayjkaiaawMcaam
% aaBaaaleaacaWGWbGaam4BaiaadYgaaeqaaOGaeyypa0ZaaeWaaeaa
% daWcaaqaaiaadsgacqGHdpWCaeaacaWGKbGaeyyQdCfaamaabmaaba
% GaeqiUdehacaGLOaGaayzkaaaacaGLOaGaayzkaaWaaSbaaSqaaiaa
% dwhacaWGUbGaamiCaiaad+gacaWGSbGaaiOlaaqabaGcdaWadaqaai
% aaigdacqGHsisldaWcaaqaaiaaigdaaeaacaaIYaaaaiabgs8a0naa
% BaaaleaacaaIYaGaaGimaaqabaGccaWGubWaaSbaaSqaaiaaikdaca
% aIWaaabeaakiabgUcaRiaadMgadaGcaaqaaiaaikdaaSqabaGccqGH
% epaDdaWgaaWcbaGaaGymaiaaicdaaeqaaOGaamivamaaBaaaleaaca
% aIXaGaaGymaaqabaGcciGGJbGaai4BaiaacohacqGHvpGAcqGHsisl
% daGcaaqaamaalaaabaGaaG4maaqaaiaaikdaaaaaleqaaOGaeyiXdq
% 3aaSbaaSqaaiaaikdacaaIWaaabeaakiaadsfadaWgaaWcbaGaaGOm
% aiaaikdaaeqaaOGaci4yaiaac+gacaGGZbGaaGOmaiabgw9aQbGaay
% 5waiaaw2faaaaa!7C7C!
\begin{equation}\label{equ:pol_X_section}
\left( {\frac{{d\sigma }} {{d\Omega }}\left( {\theta ,\varphi }
\right)} \right)_{pol}  = \left( {\frac{{d\sigma }} {{d\Omega
}}\left( \theta  \right)} \right)_{unpol.} \left[ {1 - \frac{1}
{2}\tau _{20} T_{20}  + i\sqrt 2 \tau _{10} T_{11} \cos \varphi  -
\sqrt {\frac{3} {2}} \tau _{20} T_{22} \cos 2\varphi } \right]
\end{equation}
with $\tau_{10}$ and $\tau_{20}$ the vector and tensor polarization
of the beam. The unpolarized cross section is the sum of the
amplitudes squared. The relation between the corresponding analyzing
powers $T_{ik}$ and the amplitude components is then
% MathType!MTEF!2!1!+-
% feaaeaart1ev0aaatCvAUfeBSjuyZL2yd9gzLbvyNv2CaerbuLwBLn
% hiov2DGi1BTfMBaeXatLxBI9gBaerbd9wDYLwzYbItLDharqqtubsr
% 4rNCHbGeaGqiVu0Je9sqqrpepC0xbbL8F4rqqrFfpeea0xe9Lq-Jc9
% vqaqpepm0xbba9pwe9Q8fs0-yqaqpepae9pg0FirpepeKkFr0xfr-x
% fr-xb9adbaqaaeGaciGaaiaabeqaamaabaabaaGceaqabeaacaWGub
% WaaSbaaSqaaiaaigdacaaIXaaabeaakiabg2da9maalaaabaGaaG4m
% aaqaaiaaikdadaGcaaqaaiaaigdacaaIWaaaleqaaaaakiGacMeaca
% GGTbGaaiikaiaadggadaWgaaWcbaGaaGimaaqabaGccaWGHbWaa0ba
% aSqaaiaaigdaaeaacaGGQaaaaOGaaiykaiGacohacaGGPbGaaiOBai
% abeI7aXbqaaiaadsfadaWgaaWcbaGaaGOmaiaaicdaaeqaaOGaeyyp
% a0ZaaSaaaeaacaaIXaaabaGaaG4maaaacaWGHbWaa0baaSqaaiaaic
% daaeaacaaIYaaaaOGaeyOeI0YaaSaaaeaacaaI5aaabaGaaGymaiaa
% icdaaaGaamyyamaaDaaaleaacaaIXaaabaGaaGOmaaaakiGacohaca
% GGPbGaaiOBamaaCaaaleqabaGaaGOmaaaakiabeI7aXbqaaiaadsfa
% daWgaaWcbaGaaGOmaiaaikdaaeqaaOGaeyypa0ZaaSaaaeaacaaI5a
% WaaOaaaeaacaaIZaaaleqaaaGcbaGaaGinaiaaicdaaaGaamyyamaa
% DaaaleaacaaIXaaabaGaaGOmaaaakiGacohacaGGPbGaaiOBamaaCa
% aaleqabaGaaGOmaaaakiabeI7aXbaaaa!6A61!
\begin{equation}
\begin{gathered}
  T_{11}  = \frac{3}
{{2\sqrt {10} }}\operatorname{Im} (a_0 a_1^* )\sin \theta  \hfill \\
  T_{20}  = \frac{1}
{3}a_0^2  - \frac{9}
{{10}}a_1^2 \sin ^2 \theta  \hfill \\
  T_{22}  = \frac{{9\sqrt 3 }}
{{40}}a_1^2 \sin ^2 \theta .  \hfill \\
\end{gathered}
\end{equation}
From four observables, one can then deduce the two real and two
imaginary parts of the amplitudes. The knowledge of the amplitudes
is of importance in the context of a bound state. A recent analysis
of the scattering length from $pd\to {^3He}\eta$ yielded a very
small imaginary part and uncertainty about the sign of the real part
\cite{Sibirtsev05}. This is surprising since the free pionic
inelasticity of $\eta N$ scattering is large and seems to decouple
in the case of nuclei. This  decoupling or very weak absorption was
recently attributed to a suppression of the two main inelasticity
channels \cite{Niskanen05}. This is the pion inelasticity due to the
process $\eta N\to \pi N$ and the nuclear inelasticity $\eta d\to
NN\pi$ with $d$ a quasi deuteron state.

\begin{figure}[h]
\begin{center}
\includegraphics[width=8 cm]{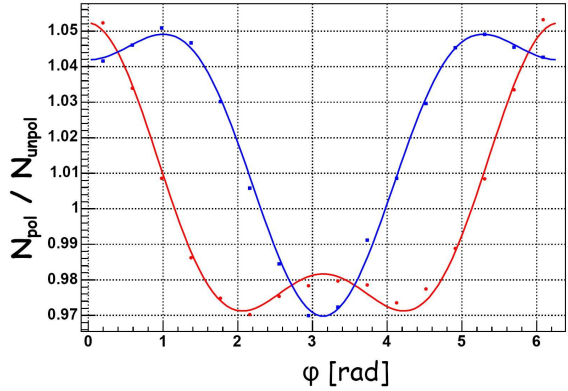}
\caption{Ratio of counting rates for polarized to unpolarized
deuteron beam as measured with the wedge detector. The curve with
the minimum at $\phi=\pi$ is for nominal $p_{zz}=-1$ the other for
$p_{zz}=+1$} \label{Fig:polarization}
\end{center}
\end{figure}
An experiment employing vector and tensor polarized deuteron beams
was performed by GEM earlier this year. The data are presently under
evaluation. In order to continuously monitor the polarization an
additional detector was mounted downstream behind the target
consisting of 16 wedge shaped scintillators. The result of such a
measurement is shown in Fig. \ref{Fig:polarization} for $p_{zz}=\pm
1$. The curves are the function $\sigma(\phi)=A(1+B\cos\varphi
+C\cos 2\varphi)$ with fitted constants $A,B$ and $C$ to the data.

Two open problems remain on the experimental side: the disagreement
between different data sets for $pd\to{^3He}\eta$ and the lack of
data for both reactions at higher beam momenta.

\section{The reaction
$p+{^6Li}\to\eta+{^7Be}$}\label{sec:reaction_li6}

\begin{figure}[h]
\begin{center}
\includegraphics[height=5 cm]{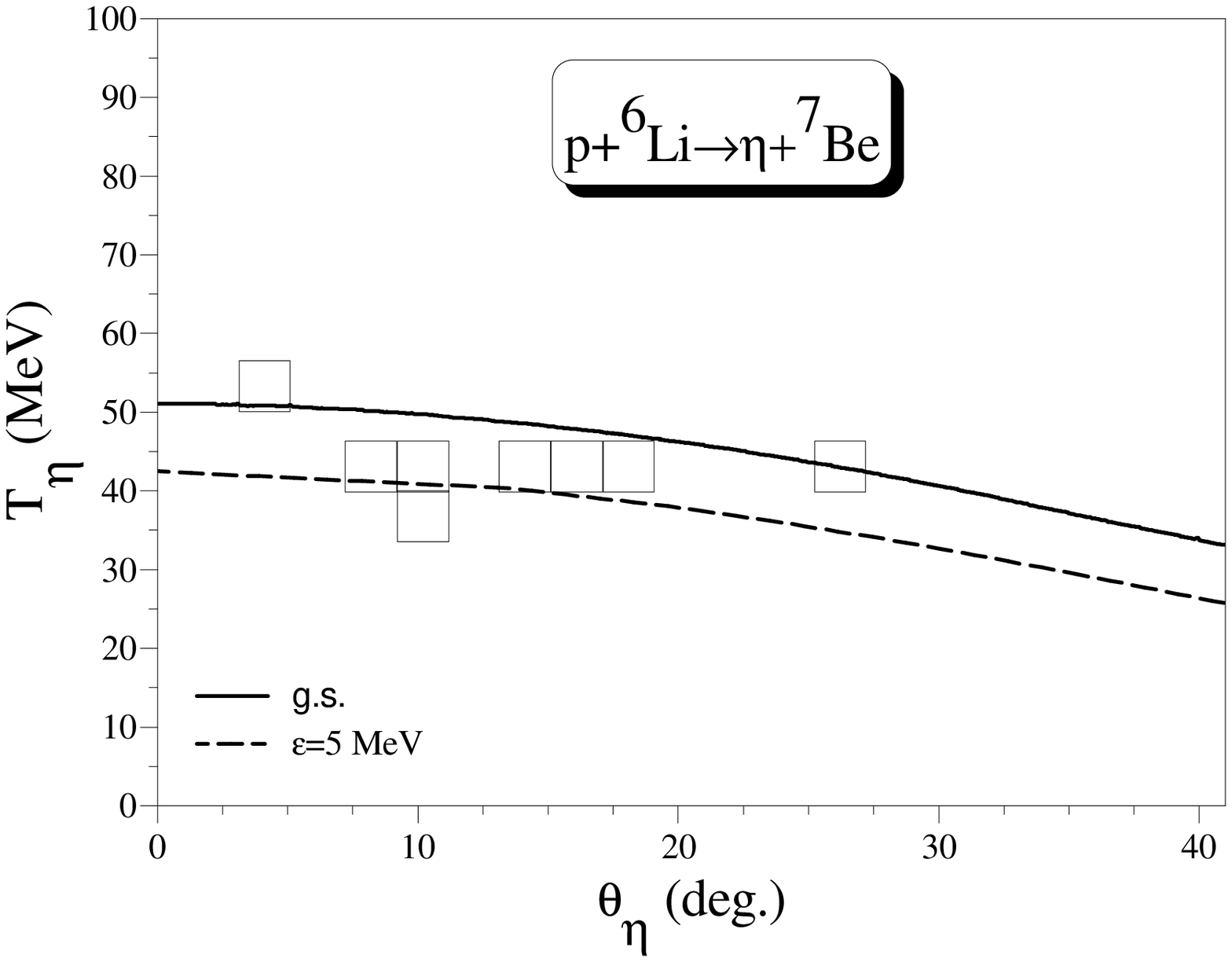}\vspace{5 mm}
\includegraphics[height=5 cm]{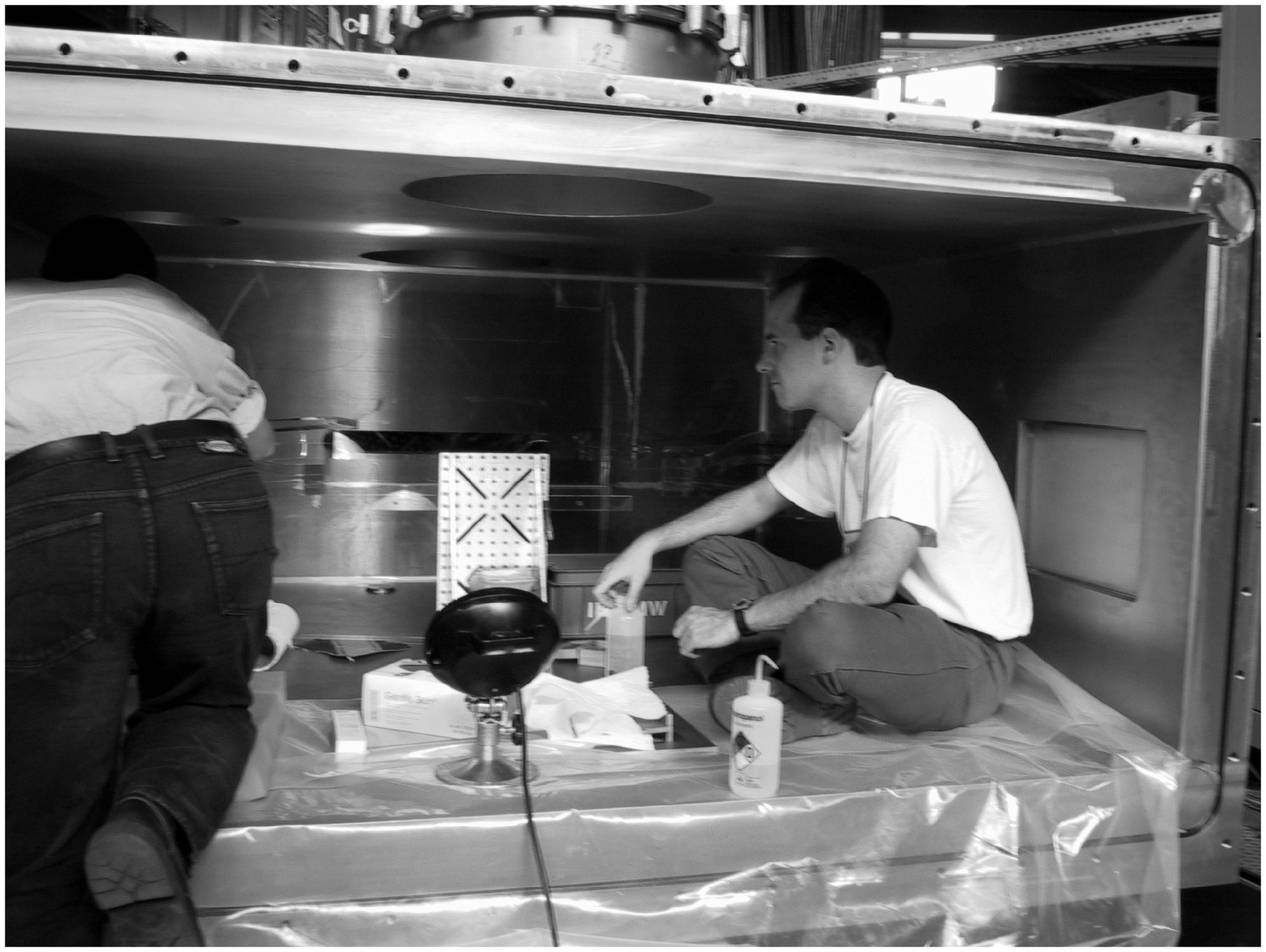}
\caption{Left: The The kinematic dependence of the data
\cite{Scomparin93}. The kinematical curves are for the ${^7Be}$
ground state and an excitation of 5 MeV. Right: Box which contains
the detectors in the focal plane in vacuum. See the slit in the
backward wall which is the entrance of the spectrograph.}
\label{Fig:Scomparin}
\end{center}
\end{figure}

This reaction has a heavier nucleus as target and thus the study of
this reaction may yield insight into the movement of a possible pole
position. The reaction was studied earlier at Saclay
\cite{Scomparin93} at a beam energy of 683 MeV. The $\eta$ was
identified via its two-photon decay. Eight events were observed.
Taking the acceptance of the detector into account a cross section
$d\sigma /d\Omega=(4.6\pm 3.8)$ nb/sr is obtained. The quoted error
is purely statistical and a systematic error of 20$\%$ should be
added. With the energy resolution of the set-up it was impossible to
distinguish different final states of the residual nucleus. Fig.
\ref{Fig:Scomparin} shows the data together with the kinematical
curve for ${^7Be}$ in its ground state and up to 5 MeV excitation.
In an accompanying theory paper it was argued that most of the yield
is from states close to 10 MeV excitation \cite{Al-Khalili93}. At
GEM the investigation of this reaction is planned at an energy
closer to threshold. In contrast to the Saclay experiment the
detection of the recoiling ${^7Be}$ with a magnetic spectrograph is
planned. The target thickness will limit the resolution to 1 MeV, So
the experiment is exclusive since all states above the first excited
state at 0.4 MeV are particle unbound. Because of the large stopping
power of the recoil ${^7Be}$ in material, (or in air,) the present
set-up of detectors in the focal plain is not adequate. All
detection elements have to operate in vacuum. For this purpose a
large vacuum box will host the detectors (see Fig.
\ref{Fig:Scomparin}).
% \begin{figure}
% \begin{center}
% \includegraphics[width=8 cm]{Box_BW.eps}
% \caption{Box which contains the detectors in the focal plane in
% vacuum. See the slit in the backward wall which is the entrance of
% the spectrograph.} \label{Fig:Box}
% \end{center}
% \end{figure}
Particle identification will be performed with a $\Delta E-E$ system
of plastic scintillators of 0.5 mm and 2 mm thickness, respectively.
Light is read out left and right by fast phototubes. In a test run
excellent particle resolution was observed, however, the position
resolution was not sufficient to perform missing mass
reconstruction. Therefore two packs of multiwire-avalanche chambers
with two dimensional position resolution have been added in front of
the $\Delta E$ counter. The run is scheduled for summer 2006.

\section{The mass of the $\eta$}\label{sec:mass-eta}

Compared to other light mesons, the mass of the $\eta$ is
surprisingly poorly known. From 1992 on the PDG ignored the old
bubble chamber data since a new measurement with an electronic
detector was published \cite{Plouin92}. Though the PDG quote in
their 2004 compilation a value of $m_{\eta}=547.75\pm
0.12~\textrm{MeV/c}^2$ \cite{PDG04}, this error hides differences of
up to 0.7~MeV/c$^2$ between the results of some of the modern
counter experiments. This new PDG average is in fact dominated by
the result of the CERN NA48 experiment,
$m_{\eta}=547.843\pm0.051~\textrm{MeV/c}^2$, which is based upon the
study of the kinematics of the six photons from the $3\pi^0$ decay
of 110~GeV $\eta$-mesons~\cite{Lai02}. In the other experiments
employing electronic detectors, which typically suggest a mass
$\approx 0.5~\textrm{MeV/c}^2$ lighter, the $\eta$ was produced much
closer to threshold and its mass primarily determined through a
missing-mass technique where, unlike the NA48 experiment, precise
knowledge of the beam momentum plays an essential part.
\begin{figure}[h]
\begin{center}
\includegraphics[width=8 cm]{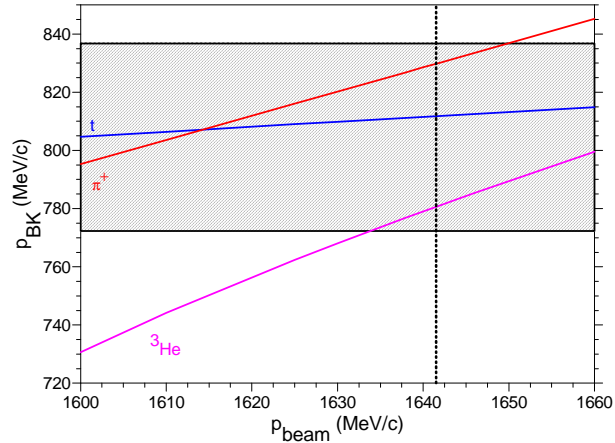}
\caption{The momenta  of the three particles of interest under zero
degree in the laboratory system as function of the beam momentum.
The acceptance of the spectrograph for a central momentum of 804.5
MeV/c is shown as shaded area and the beam momentum of 1641.4 MeV/c
is indicated.} \label{Fig:BK_acceptance}
\end{center}
\end{figure}
The Big Karl spectrograph and the high brilliance beam at COSY are
ideally suited to perform a high precision experiment. The
underlying idea of the study is a self-calibrating experiment. Three
reaction products were measured simultaneously with one setting of
the spectrometer and one setting of the beam momentum. The reaction
products were

% MathType!MTEF!2!1!+-
% feaaeaart1ev0aaatCvAUfeBSjuyZL2yd9gzLbvyNv2CaerbuLwBLn
% hiov2DGi1BTfMBaeXatLxBI9gBaerbd9wDYLwzYbItLDharqqr1ngB
% PrgifHhDYfgasaacH8srps0lbbf9q8WrFfeuY-Hhbbf9v8qqaqFr0x
% c9pk0xbba9q8WqFfea0-yr0RYxir-Jbba9q8aq0-yq-He9q8qqQ8fr
% Fve9Fve9Ff0dmeaabaqaciGacaGaaeqabaWaaeaaeaaakeaaqaa6du
% FaOpWdbuaabeqadeaaaeaaaeaacaWGWbGaey4kaSIaamizaaqaaaaa
% cqGHsgIRdaGabaqaauaabeqadeaaaeaadaahbaWcbeqaaiaaiodaaa
% GccaWGibGaey4kaSIaeqiWda3aaWbaaSqabeaacqGHRaWkaaaakeaa
% cqaHapaCdaahaaWcbeqaaiabgUcaRaaakiabgUcaRmaaCeaaleqaba
% GaaG4maaaakiaadIeaaeaadaahbaWcbeqaaiaaiodaaaGccaWGibGa
% amyzaiabgUcaRiabeE7aObaaaiaawUhaaaaa!509D!
\begin{equation}
\begin{array}{*{20}c}
   {}  \\
   {p + d}  \\
   {}  \\
\end{array}  \to \left\{ {\begin{array}{*{20}c}
   {{}^3H + \pi ^ +  }  \\
   {\pi ^ +   + {}^3H}  \\
   {{}^3He + \eta }  \\
\end{array} } \right.
\end{equation}
where it is always the third particle which will be detected. The
method relies on the fact that the masses of the proton, deuteron,
$\pi^+$, triton and ${^3He}$ are well known. Fig.
\ref{Fig:BK_acceptance} shows the momenta of the third particle
being emitted at zero degree in the laboratory system. Pions and
${^3He}$ are emitted in the forward direction, tritons in the
backward direction in the center of mass system. For ${^3He}$ the
momenta were divided by two in order to account for the double
charge. The momentum acceptance of the spectrograph is also shown.
Clearly, for a beam momentum close to 1641 MeV/c all three particles
are within the acceptance of the spectrograph. The pion is used to
deduce the absolute beam momentum. Then the triton will be used to
fix the spectrograph setting and finally from the ${^3He}$ one
obtains the mass of the $\eta$ meson.

In order to fix the properties of the spectrograph a series of
calibration runs were performed. These include sweeping the primary
beam over the focal plane without a target at a beam momentum of 793
MeV/c. This corresponds to a reaction $p+0\to p+0$. Then the full
kinematical ellipse of deuterons from the reaction $p+p\to d+\pi^+$
at the same beam momentum was measured. Finally, pions from the
reaction $p+p\to \pi^++d$ were measured at $\approx$ 1640 MeV/c
while again sweeping the deuteron loci over the whole focal plane.
\begin{figure}[t]
\begin{center}
\includegraphics[width=8 cm]{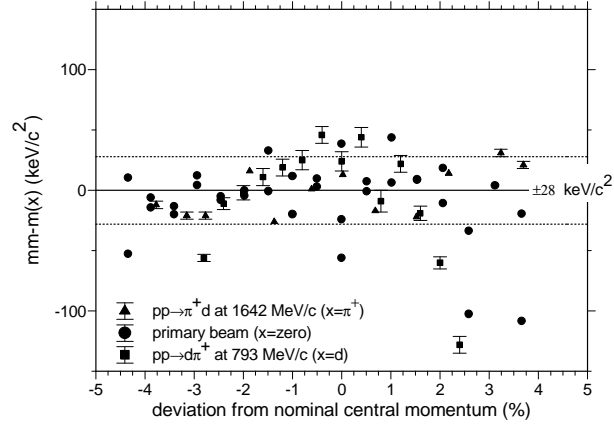}
\caption{The deviation of the measured missing mass from its PDG
value as function of the deviation from the nominal central value
for the momentum of the magnetic spectrograph.}
\label{Fig:mm_deviation}
\end{center}
\end{figure}
\begin{figure}[h]
\begin{center}
\includegraphics[width=8 cm]{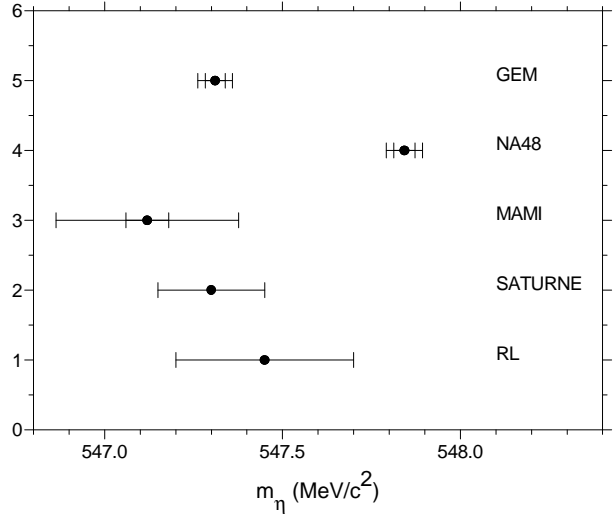}
\caption{The results of the $\eta$-mass measurements, in order of
publication date, taken from the Rutherford Laboratory
(RL)~\cite{Duane74}, SATURNE~\cite{Plouin92}, MAMI~\cite{Krusche95},
NA48~\cite{Lai02}, and GEM. When two error bars are shown, the
smaller is statistical and the larger total.}
\label{Fig:mass_comparison}
\end{center}
\end{figure}
In the analysis the target thickness as measured from the triton
momentum was studied as function of measuring time. It was found
that it increased with time most probably due to freezing out of
air. Evidence for a thinner target at the midpoint of the
measurement corresponds with a cleaning of the target windows. Then
the following procedure was adopted. It is assumed that the
properties of the spectrograph are known. The three calibration
reactions were now used to fix the beam momentum, the target
thickness and the $\eta $ mass. In a second step the assumption
(known spectrograph) was examined by determining the missing mass of
the unobserved particle in the calibration runs. These are the
masses 0, $\pi^+$ and $d$. The result of this exercise is shown in
Fig. \ref{Fig:mm_deviation}. It shows the deviation of the measured
missing mass from its nominal value \cite{PDG04} as function of the
relative momentum difference of the central momentum setting of the
spectrograph. It can be inferred that the deviation has an
uncertainty of $\sigma=\pm 28$ keV/c$^2$ which is the main
contribution to the systematic error which in total is 32 keV/c$^2$.
The missing mass measurement yields a statistical error of the same
order of magnitude. The final result is \cite{Abdel-Bary05}:
\begin{equation}
m(\eta)=547.311\pm 0.028\text{ (stat.) }\pm 0.032\text{ (syst.)
}\text{MeV/c}^2.
\end{equation}
Finally this number is compared with the other values presently
recognized by the PDG \cite{PDG04} (see Fig.
\ref{Fig:mass_comparison}). The present mass is in agreement with
the earlier results employing $\eta$ production. It disagrees with
the value from $\eta$ decay.
\section{Summary}
GEM has measured a series of differential as well as total cross
sections for proton and deuteron  projectiles and light nuclei as
targets. Obviously the interaction in the final state in these cases
is very different from the $\eta N$ interaction. The reason might be
that the elementary scattering length does not have the same isospin
algebraic and spatial properties for the real and imaginary part as
in a nuclear medium. A positive value of the real part means a
modest attraction while a negative value means repulsion or a bound
state. A small imaginary part in the case of nuclei should lead to a
more narrow state if a bound $\eta$-nuclear state exists
\cite{Sibirtsev04}. The experiments will allow the determination of
size and signs of the scattering length components. A dedicated
search for a possible bound $\eta$-nuclear state has started. Basic
properties of the $\eta$ such as its mass and width were poorly
known. A new value for the mass has been derived with extremely
small error bars. The width is only known up to 50$\%$ \cite{PDG04}
and needs further measurements.

%GATHER{d:\texmf\bibtex\bib\eigene\meson.bib}
%\bibliography{meson}

\end{document}